\documentclass[prl,twocolumn,superscriptaddress]{revtex4-1}
\pdfoutput=1
\usepackage[letterpaper,top=1.45cm,bottom=1.65cm,left=2cm,right=2cm]{geometry}
\usepackage{amssymb,amsmath,amsthm,graphicx,mathptmx}
\usepackage{subfigure}
\usepackage{graphics, color}
\usepackage{latexsym}
\usepackage{bm}
\usepackage{epsfig}
\usepackage{multirow,tabularx}
\usepackage[colorlinks=true,linktocpage=true,linkcolor=blue,citecolor=blue]{hyperref}
\usepackage[abs]{overpic}

\usepackage{fancyhdr}
\fancypagestyle{plain}

\newcommand{\PRLsection}[1]{\emph{#1.---}}

\newcolumntype{Y}{>{\centering\arraybackslash}X}

\begin{document}

\title{Non-Spherically Symmetric Collapse in Asymptotically AdS Spacetimes}

\author{Hans Bantilan}
\email{h.bantilan@qmul.ac.uk}
\affiliation{School of Mathematical Sciences, Queen Mary University of London, Mile End Road, London E1 4NS, United Kingdom}
\affiliation{Centre for Research in String Theory, School of Physics and Astronomy, Queen Mary University of London, E1 4NS, UK}
\affiliation{Department of Applied Mathematics and Theoretical Physics (DAMTP), Centre for Mathematical Sciences, University of Cambridge, Wilberforce Road, Cambridge CB3 0WA, United Kingdom}
\author{Pau Figueras}
\email{p.figueras@qmul.ac.uk}
\affiliation{School of Mathematical Sciences, Queen Mary University of London, Mile End Road, London E1 4NS, United Kingdom}
\author{Markus Kunesch}
\email{m.kunesch@damtp.cam.ac.uk}
\affiliation{Department of Applied Mathematics and Theoretical Physics (DAMTP), Centre for Mathematical Sciences, University of Cambridge, Wilberforce Road, Cambridge CB3 0WA, United Kingdom}
\author{Paul Romatschke}
\email{paul.romatschke@colorado.edu}
\affiliation{Department of Physics, University of Colorado, Boulder, Colorado 80309, USA}
\affiliation{Center for Theory of Quantum Matter, University of Colorado, Boulder, Colorado 80309, USA}

\begin{abstract}
We numerically simulate gravitational collapse in asymptotically anti-de Sitter spacetimes away from spherical symmetry. Starting from initial data sourced by a massless real scalar field, we solve the Einstein equations with a negative cosmological constant in five spacetime dimensions and obtain a family of non-spherically symmetric solutions, including those that form two distinct black holes on the axis. We find that these configurations collapse faster than spherically symmetric ones of the same mass and radial compactness. Similarly, they require less mass to collapse within a fixed time.
\end{abstract}

\maketitle

%-------------------------------------------------------
% Introduction
%-------------------------------------------------------
\PRLsection{Introduction}%
The effort to understand the dynamics of gravity in asymptotically anti-de Sitter (AdS) spacetimes 
is driven by a series of open questions that has generated intense interest in recent years.
As the maximally symmetric solution of the Einstein equations with a negative cosmological constant 
$(\Lambda<0)$, AdS space is as fundamental as Minkowski space $(\Lambda=0)$, the non-linear 
stability of which was established by the work of Ref. \cite{Christodoulou:1993uv}. 
Similar stability results for de Sitter space $(\Lambda>0)$ were proven in 
\cite{Friedrich1986,FRIEDRICH1995125}. 
A question related to stability is that of black hole formation: how do black holes form in these
spacetimes? 
The answer to this question has implications for several open problems in general relativity, 
including the validity of the weak cosmic censorship conjecture \cite{Penrose:1969pc}.
Black hole formation in asymptotically flat spacetimes was studied mathematically by
Ref. \cite{Christodoulou:1987vv,Christodoulou:1991yfa} in spherical symmetry, and by
Ref. \cite{Choptuik:1992jv} using numerical techniques that led to the discovery of the celebrated
critical phenomena in gravitational collapse; see Ref. \cite{Jalmuzna:2017mpa} for a recent study.
In the asymptotically flat setting, considerable progress has been made with no symmetry
assumptions by Ref. \cite{Christodoulou:2008nj}; see also Ref. 
\cite{Klainerman:2009ga,Klainerman:2013tga}.

The dynamics of gravity in asymptotically AdS spacetimes is not as well-understood as the 
asymptotically flat case.
One reason for this is that when $\Lambda<0$, solving the Einstein equations constitutes an initial
boundary value problem.
In contrast to the asymptotically flat case, the boundary of AdS is timelike and in causal contact 
with its interior.
For the reflecting boundary conditions that are commonly used in the literature, 
Ref. \cite{Dafermos1,Dafermos2} conjectured that AdS is non-linearly unstable to the formation of 
black holes; see also Ref. \cite{Anderson:2006ax}. 
In spherical symmetry, Ref. \cite{Moschidis:2017lcr,Moschidis:2017llu} provided a proof for this 
instability in the specific setting of the Einstein-null dust system with an inner mirror. 
Black hole formation in AdS acquires an added significance in light of the AdS/CFT 
correspondence \cite{Maldacena:1997re,Gubser:1998bc,Witten:1998qj}, according to which black hole 
formation in AdS corresponds to thermalization in a dual conformal field theory (CFT).
The question of whether or not a black hole generically forms in AdS is then related to the 
question of whether or not a state in a strongly interacting CFT generically thermalizes.

Gravitational collapse in AdS was first studied in three dimensions in Ref. 
\cite{Pretorius:2000yu}. 
Asymptotically AdS spacetimes in $D$ spacetime dimensions (AdS$_D$) are special for $D=3$: there is 
a mass gap, so configurations with a total mass below a certain threshold cannot undergo 
gravitational collapse
\footnote{This was revisited by Ref. \cite{Jalmuzna:2017mpa} that studied critical scalar collapse 
with a rotating complex scalar field in AdS$_3$ spacetimes, where a rotational Killing vector 
reduces the evolution to a $1+1$ dimensional problem.}, even though such solutions may not remain 
close to AdS$_3$ in any reasonable norm \cite{Bizon:2013xha}.
For $D \ge 4$, configurations that collapse into a black hole after several bounces from the AdS 
boundary were found in Ref. \cite{Bizon:2011gg}; see also Ref. 
\cite{Liebling:2012gv,Buchel:2012uh}. 
In those studies, spherical symmetry was imposed, and the dynamics was driven by the presence of a 
massless scalar field. 
Initial data with an amplitude $\epsilon$ were numerically evolved in time and were found to 
develop features at progressively smaller spatial scales, in a turbulent process that terminates in 
gravitational collapse on a timescale of $O(\epsilon^{-2})$. 
On the other hand, within the same model in spherical symmetry, it was also found that there exist 
open sets of initial data that lead to solutions that are non-linearly stable against 
gravitational collapse \cite{Dias:2012tq,Maliborski:2013jca,Balasubramanian:2014cja}. 

Efforts have begun to extend these studies beyond spherical symmetry 
\cite{Horowitz:2014hja,Dias:2016ewl,Rostworowski:2017tcx,Martinon:2017uyo,Dias:2017tjg}.
The majority of results to date have been perturbative in nature
\footnote{The notable exceptions are the numerical construction of time-periodic AdS geon solutions 
in Ref. \cite{Horowitz:2014hja,Martinon:2017uyo}; these solutions never collapse, by 
construction.}, and as such, cannot directly address the question of whether or not black holes 
form. 
In this Letter, we present the first study of gravitational collapse in AdS with inhomogeneous 
deformations away from spherical symmetry.
For the numerical simulations presented here, we specialize to the case of global AdS$_5$, and we 
address the following question:  
``{\em does gravitational collapse in AdS occur earlier or later away from spherical symmetry?}'' 
We answer this question by constructing fully non-linear time-dependent solutions of the Einstein 
equations starting with initial data sourced by a massless real scalar field. 

%-------------------------------------------------------
% Numerical Scheme
%-------------------------------------------------------
\PRLsection{Numerical Scheme}%
The results presented in this Letter are based on a new numerical code to solve the Einstein 
equations for asymptotically AdS spacetimes. 
This code uses Cartesian coordinates in global AdS and is based on generalized harmonic evolution 
\cite{Pretorius:2004jg}; see also Ref. \cite{Bantilan:2012vu,Bantilan:2014sra}.  
We define Cartesian coordinates in the following way: consider the metric of global AdS$_5$,
\begin{equation}
\hat g =  -\left( 1+\frac{r^2}{L^2}\right) dt^2 + \frac{dr^2}{1+\frac{r^2}{L^2}}  +r^2 d\Omega_{(3)}^2,
\label{eqn:pure_ads_metric}
\end{equation}
where $L$ is the AdS radius, which is related to the cosmological constant by $\Lambda=-6/L^2$, and 
$d\Omega_{(3)}^2 = d\chi^2 + \sin^2\chi (d\theta^2 + \sin^2\theta d\phi^2)$ is the metric on the 
unit round 3-sphere.
We compactify the radial coordinate by defining $r=2\rho/(1-\rho^2/\ell^2)$ so that the AdS 
boundary, $\rho=\ell$, is included in the computational domain. 
Here, $\ell$ is an arbitrary compactification scale, independent of the AdS radius $L$.

In polar coordinates, the Courant-Friedrichs-Lewy (CFL) condition at the origin $\rho=0$ imposes a 
severe restriction on the size of the time step. 
We bypass this issue by introducing Cartesian coordinates $x=\rho \cos \chi$ and 
$y=\rho \sin \chi$. 
Setting $L=1$ and $\ell=1$, the metric \eqref{eqn:pure_ads_metric} in these Cartesian coordinates 
becomes
\begin{eqnarray}
\hat{g} 
&=& \frac{1}{(1-\rho^2)^2} \left[ -\hat{f}(\rho) dt^2 + 4 (dx^2 + dy^2 + y^2 d\Omega_{(2)}^2) \right]\,, 
\end{eqnarray}
where $\hat{f}(\rho) = (1-\rho^2)^2+4\rho^2$, and 
$d\Omega_{(2)}^2 = d\theta^2 + \sin^2\theta d\phi^2$ is the metric on the unit round 2-sphere.

In moving away from pure AdS, we preserve an $SO(3)$ symmetry that acts to rotate the 2-spheres 
parametrized by $\theta$ and $\phi$. 
This implies that there are seven independent metric components 
$g_{tt}, g_{tx}, g_{ty}, g_{xx}, g_{xy}, g_{yy}, g_{\theta\theta}$, each of which depends on 
$(t,x,y)$.  
With this $SO(3)$ symmetry in five dimensions, the general form of the full metric away from pure 
AdS is
\begin{equation}
\begin{aligned}
g 
=&~g_{tt}\, dt^2 + g_{xx}\, dx^2 + g_{yy}\, dy^2 + g_{\theta\theta}\, d\Omega_{(2)}^2  \\
& + 2\, \left( g_{tx}\, dt \,dx + g_{ty} \,dt\, dy + g_{xy} \,dx \,dy \right).
\end{aligned}
\end{equation}
In our Cartesian coordinates, the axis of symmetry where the 2-sphere shrinks to zero size is at 
$y=0$. 
To ensure that the spacetime remains smooth there, we impose suitable regularity conditions on our 
evolved variables at those points. See Supplemental Material for details of our evolution scheme together with the gauge choice, boundary conditions, and axis regularity conditions \cite{sm_reference}.

\begin{figure}[t]
         \includegraphics[width=\linewidth]{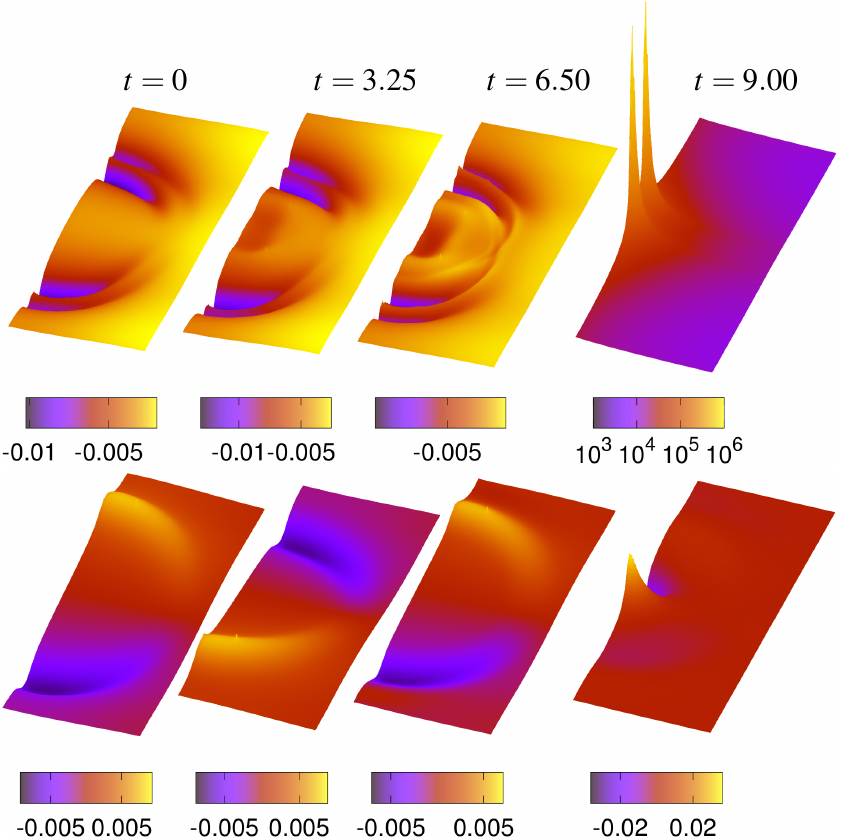}
                        {\caption{
                          \textit{Top}: Snapshots of $K/K_{AdS}-1$ in global AdS at different global times $t$ in units of $L$ for an initial profile $A=0$, $B=0.0087$ and a total mass $M=0.021$ in units of $L^2$. The evolution is quasi-periodic until the first AHs are detected at $t=9.04$. 
                          \textit{Below}: Snapshots of the scalar field profile for the same configuration, with $\rho_c=0.4$ and $\rho_d=0.8$. For each of the panels, the top and bottom edges are $x=-1,1$ and the left and right edges are $y=0,1$ respectively. 
                        }\label{fig:gfn}}
\end{figure}

We couple gravity to a massless real scalar field $\varphi$, and construct time-symmetric data on 
the initial time slice by solving the Hamiltonian constraint subject to a freely chosen initial 
scalar field profile. 
We use a family of profiles that smoothly interpolate between spherically symmetric and 
non-spherically symmetric configurations
\begin{equation}\label{eqn:scalar_profile}
\varphi(\rho,\chi) = A f(\rho) + B\, g(\rho) \cos\chi,
\end{equation}
where $f(\rho)$ and $g(\rho)$ are $C^2$ transition functions that spatially vary in the range  
$\rho\in\left[\rho_c,\rho_d \right]$, for some arbitrary $\rho_c$, $\rho_d$; 
see Fig.~\ref{fig:gfn}. 
The spatial gradients of these data are compactly supported in the radial direction in the shape of 
an annulus centered around $\rho_0 \in [\rho_c,\rho_d]$. The constants $A$ and $B$ measure the 
strength of the spherically symmetric and the non-spherically symmetric terms of the initial data 
respectively. 
Our results are not sensitive to the specific choice of $f(\rho)$ and $g(\rho)$. 
See Supplemental Material for the details of the typical transition functions that we use in our 
simulations.
Given our choice of initial data, the total angular momentum of the spacetime is zero. 
Therefore, if weak cosmic censorship holds in our setup, the final state of each of our simulations 
should be a global Schwarzschild-AdS black hole.

We monitor the evolution of black holes by keeping track of trapped surfaces.
We excise a portion of the interior of any apparent horizon (AH) that forms, to remove any 
singularities from the computational domain. 
No boundary conditions are imposed on the excision surface; instead, the Einstein equations are 
solved there using one-sided stencils.
We use Kreiss-Oliger dissipation \cite{kreiss1973methods} to damp unphysical high-frequency modes 
that can arise at grid boundaries, with a typical dissipation parameter of $0.35$. 

We numerically solve the Einstein equations and the Hamiltonian constraint using the PAMR/AMRD 
libraries \cite{PAMR}, and discretize the equations using second order finite differences.
The evolution equations for the metric and scalar field are integrated in time using an iterative 
Newton-Gauss-Seidel relaxation procedure. 
The numerical grid is in $(t,x,y)$ with $t \in [0,t_{max}]$, $x \in [-1,1]$, $y \in [0,1]$. 
A typical \hbox{unigrid} resolution has $N_x=1025$, $N_y=513$ grid points with equal grid spacings 
$\Delta x=\Delta y$ in the Cartesian directions. 
We use a typical Courant factor of $\lambda \equiv \Delta t /\Delta x = 0.2$. 
The results presented here were obtained with \hbox{unigrid} or with fixed refinement although the 
code has adaptive mesh refinement capabilities. 
See Supplemental Material for convergence tests.

%-------------------------------------------------------
% Results
%-------------------------------------------------------
\PRLsection{Results}%
For a fixed total mass and radial compactness, the time evolution of initial configurations 
\eqref{eqn:scalar_profile} further away from spherical symmetry consistently exhibit gravitational 
collapse within fewer bounces.
Throughout, we choose initial data where the spatial gradients of the initial scalar field profile 
are non-vanishing in an annular region bounded by $\rho_c=0.4$ and $\rho_d=0.8$.  
These configurations form black holes after a variable number of bounces that depend on the values 
of the coefficients $A$ and $B$ in \eqref{eqn:scalar_profile}.
The size and location of the first AH to form also depend on the values of $A$ and $B$. 
For the special case of $B=0$, i.e., spherically symmetric data, gravitational collapse leads 
directly to the formation of a black hole centered at the origin $x=y=0$. 
Non-zero $B$ cases correspond to a one-parameter family of non-spherically symmetric 
configurations sourced by a scalar field whose initial profile has a $\cos\chi$ dependence. 

Fig.~\ref{fig:gfn} (\textit{top}) shows several snapshots of the normalized difference 
$K/K_{AdS}-1$ between the Kretschmann scalar $K=R_{\alpha\beta\rho\sigma}R^{\alpha\beta\rho\sigma}$ 
and its pure AdS value $K_{AdS}$ for one such configuration with $A=0$, $B=0.0087$ and a total mass 
$M=0.021$ in units of $L^2$. 
Fig.~\ref{fig:gfn} (\textit{bottom}) shows the corresponding profiles of the scalar field. 
These data collapse after two bounces into two distinct black holes centered at antipodal points 
$x=\pm 0.12$ on the axis $y=0$. 
Aside from the last snapshot, which corresponds to a time slice shortly before collapse at 
$t=9.04$ in units of $L$, the snapshots are taken after each bounce to emphasize the 
quasi-periodic nature of the evolution. 
Deviations away from strictly periodic behavior are evidenced by a gradual sharpening of spatial 
gradients after each bounce, which eventually leads to the formation of two black holes on the 
axis. 
See Supplemental Material for the energy density in the dual conformal field theory on $S^3$ which 
provides another view of this evolution.

\begin{figure*}[t]
  \includegraphics[width=0.48\linewidth]{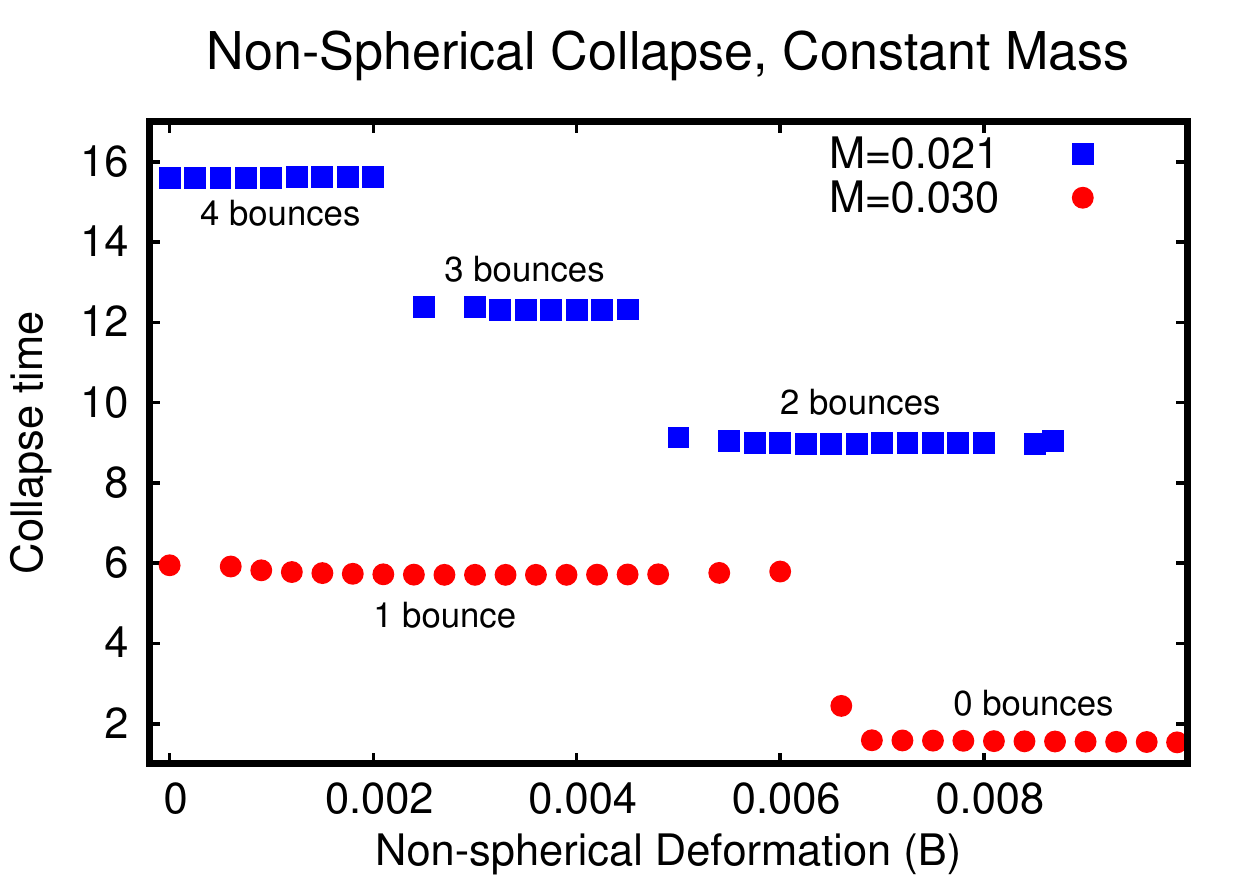}
  \hfill
  \includegraphics[width=0.48\linewidth]{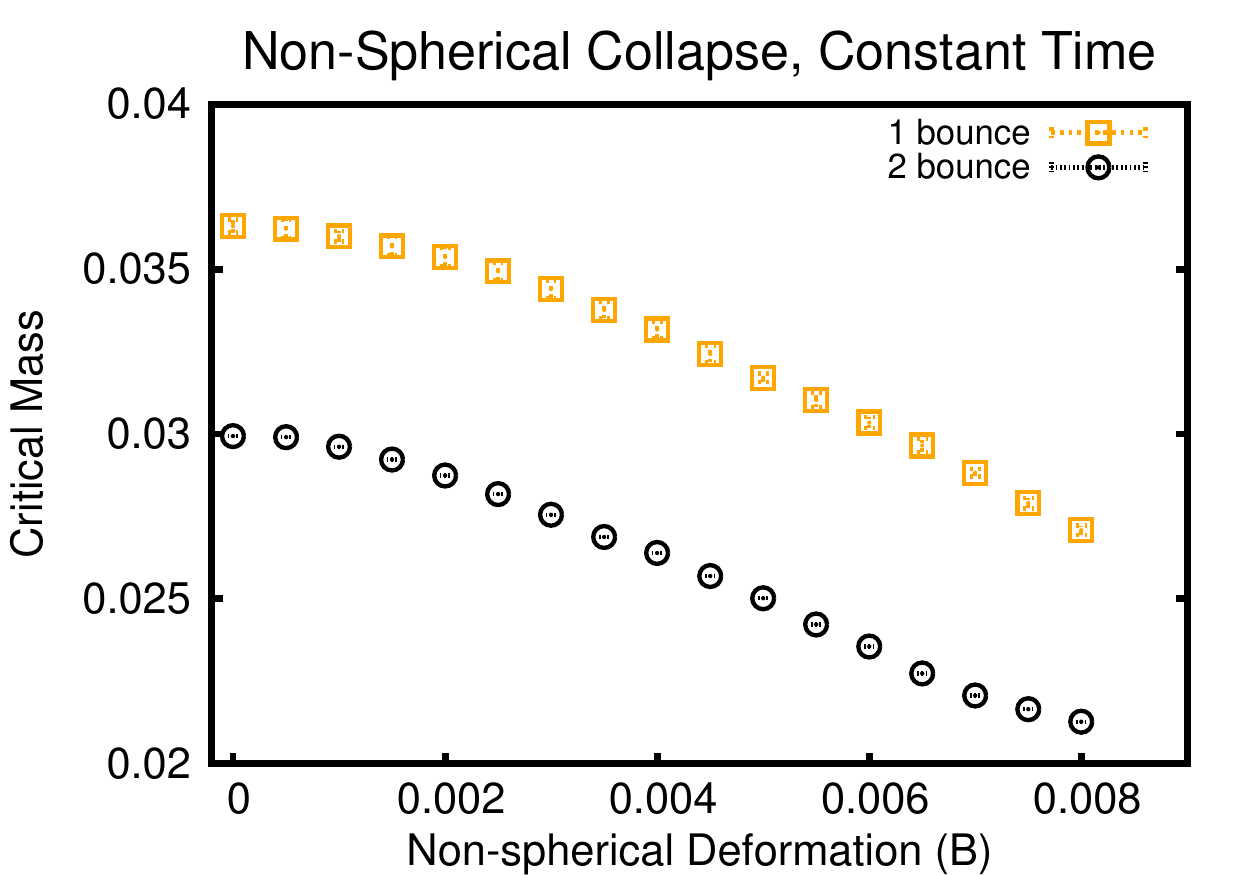}
                        {\caption{
                          \textit{Left}: Collapse time versus non-spherically symmetric deformation amplitude $B$ for fixed total masses $M=0.021$ (blue squares) and $M=0.030$ (red circles). As the value of $B$ is increased, the total mass is kept constant by decreasing the value of $A$. Non-spherically symmetric configurations form black holes earlier, i.e., in fewer bounces. The collapse time is a discontinuous function of the deformation parameter $B$.
                          \textit{Right}: Critical mass versus non-spherically symmetric deformation $B$. The critical mass is defined as the maximum mass for which a black hole is formed after $N$ bounces, shown here for $N=1$ (yellow squares) and $N=2$ (black circles); if the mass were increased further, a black hole would be formed one bounce earlier. For fixed $N$, configurations with larger 
$B$ exhibit lower critical masses, i.e., configurations further away from spherical symmetry require less mass to collapse within a given collapse time.
                        }\label{fig:data}}
\end{figure*}

Fig.~\ref{fig:data} depicts the effect of moving away from spherical symmetry in two complementary 
ways. 
In Fig.~\ref{fig:data} (\textit{left}), we consider a one-parameter family of initial data 
obtained by varying $A$ and $B$ in \eqref{eqn:scalar_profile} while keeping the total mass of the 
spacetime fixed for two representative cases with total masses $M=0.021$ (blue squares) and 
$M=0.030$ (red circles). 
The left-most point for each mass case corresponds to a spherically symmetric initial configuration 
with $B=0$. 
For these spherically symmetric points, collapse occurs in the manner that had been found in 
Ref. \cite{Bizon:2011gg}. 
Data obtained for $M=0.023$, $M=0.028$, and $M=0.084$ show similar qualitative behavior.
Increasing $B$ in \eqref{eqn:scalar_profile} has the effect of deforming the initial data away from 
spherical symmetry. 
As $B$ is increased, $A$ is decreased in order to keep the total mass fixed. 
For larger $B$, collapse occurs earlier: at certain critical values of $B$, the collapse time 
decreases by roughly $\pi L$, i.e., the two AdS light-crossing times that it takes for a bounce. 
The center of collapse also shifts further away from the origin as $B$ is increased.   
The coefficient $B$ eventually reaches a maximum value corresponding to a maximally non-spherically 
symmetric initial configuration $(A=0)$. 
These appear in Fig.~\ref{fig:data} (\textit{left}) as the right-most points. 
For these points, the initial data collapse into two distinct black holes on the axis, in fewer 
bounces than it takes their spherically symmetric counterparts to collapse into a single black 
hole at the origin.

Fig.~\ref{fig:data} (\textit{right}) depicts a different way of visualizing this faster collapse as 
one moves away from spherical symmetry.
Here, the values for $A$ and $B$ are obtained in such a way that the data are at the cusp of 
collapse after one bounce (yellow squares) and at the cusp of collapse after two bounces 
(black circles), i.e., increasing either $A$ or $B$ would result in collapse one bounce 
earlier. 
In practice, this entails increasing $B$ while keeping $A$ fixed, thereby increasing the 
total mass of the configuration, until one finds the value for $B$ where collapse time decreases 
by $\pi L$. 
Fig.~\ref{fig:data} (\textit{right}) shows that for configurations with larger $B$, 
less mass is required to stay at the cusp. 
Hence, for the family \eqref{eqn:scalar_profile} of initial profiles, non-spherically 
symmetric configurations require less mass to collapse within a given collapse time than
their spherically symmetric counterparts.

\begin{figure}[t]
  \includegraphics[width=\linewidth]{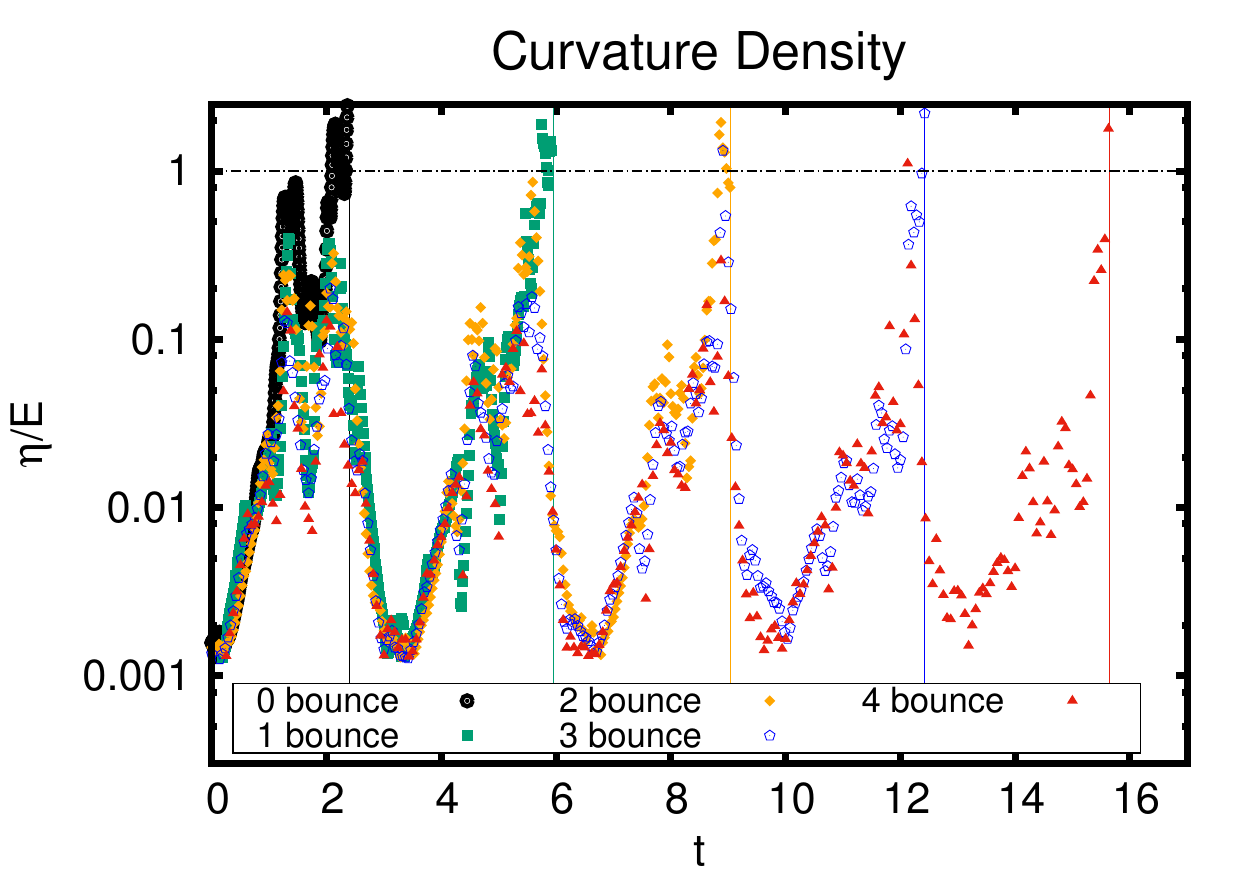}
                        {\caption{ 
                            The dimensionless ratio $\eta/E$ defined in \eqref{eqn:eta} for various representative cases that collapse after zero, one, two, three, or four bounces. For each case, this dimensionless number begins to exceed unity within a bounce prior to the formation of a trapped surface. The collapse time for each case is indicated by a vertical line.
                        }\label{fig:etaE}}
\end{figure}

Gravitational collapse is preceded by the appearance of large curvatures near the axis. 
This provides a way to anticipate when collapse will occur, even before the first trapped surface 
is detected. 
We quantify the curvature deformation of any given time slice away from pure AdS in the following 
way. 
For some arbitrary threshold value $\delta$, consider the spatial region $\mathcal{R}$ where 
$|K/K_{AdS}-1| > \delta$ on a given time slice with intrinsic metric $\gamma$. 
Construct the quantity 
\begin{equation}\label{eqn:eta}
\eta \equiv \int_{\mathcal{R}} d^4x \sqrt{\det\gamma} \left| K/K_{AdS} - 1 \right|\,,
\end{equation}
and compare it to the spatial volume $E \equiv \int_{\mathcal{R}} d^4x\sqrt{\det\gamma}$ of the 
region $\mathcal{R}$.
Fig.~\ref{fig:etaE} shows how the dimensionless number $\eta/E$ behaves over time for various 
representative cases with different collapse times. 
The oscillations in $\eta/E$ are inherited from the quasi-periodic evolution that is depicted in 
Fig~\ref{fig:gfn}. 
In all cases, trapped surfaces are formed within a bounce of $\eta/E$ exceeding roughly unity.  
We have begun to extend this study to initial data where the deformation away from spherical 
symmetry is parametrized not just in terms of the lowest spherical harmonic $\sim \cos\chi$ on 
$S^3$, but also in terms of the higher harmonics.
In these cases, $\eta/E$ continues to be a useful quantity to signal collapse, i.e., collapse is 
preceded by a time where $\eta/E \gtrsim 1$. 
It is also important to note that the time at which $\eta/E$ peaks is robust under changes in the 
threshold $\delta$, even though the actual value of $\eta/E$ that precedes collapse does depend on 
$\delta$.
See Supplemental Material for the data corresponding to these statements.

Configurations of the form \eqref{eqn:scalar_profile} with $A=0$ and $B \neq 0$ generically 
collapse into two black holes on the axis which subsequently merge to form a single 
Schwarzschild-AdS black hole. 
The time it takes to reach this final state corresponds to the thermalization time of the dual CFT. 
The discrepancy between thermalization time and collapse time is therefore the time it takes for 
the two black holes to merge and ring down to a Schwarzschild-AdS black hole. 
This discrepancy is bounded: the merger time is bounded from above by $(\pi/2) L$, i.e., the length 
of a timelike geodesic joining a given point to the origin, while the ring-down is exponential and 
thus takes a negligibly short period of time. 
There is a potential caveat however: Ref. \cite{Holzegel:2011uu} conjectured that 
Schwarzschild-AdS, or more generally Kerr-AdS, is dynamically unstable for generic perturbations. 
Away from spherical symmetry, it is not known whether the system equilibrates to a stable black 
hole. 
Therefore, it is possible that the dual CFT may not thermalize at all 
\footnote{Ref. \cite{Dias:2012tq} argued that Schwarzschild-AdS should remain stable for long 
times for sufficiently regular perturbations.}. 

%-------------------------------------------------------
% Discussion
%-------------------------------------------------------
\PRLsection{Discussion}%
We have presented the first study of gravitational collapse in AdS with inhomogeneous deformations 
away from spherical symmetry. 
Our results show that the evolution towards smaller spatial scales that leads to collapse persists 
in this setting, and in fact, we find that moving away from spherical symmetry {\em facilitates} 
collapse for a particular family of deformations.
For a fixed total mass and radial compactness, these non-spherically symmetric 
configurations collapse earlier than their spherically symmetric counterparts, and for fixed 
radial compactness and collapse time, they require less mass to collapse within the same number of 
bounces. 
This faster collapse may have already been guessed from the smaller spatial scales that are 
present in the initial profiles to begin with.
For the annular configurations that we construct, breaking spherical symmetry amounts to 
introducing spatial gradients along the annulus. 
Nevertheless, it was far from clear whether faster collapse as was observed would result from full
non-linear evolution, especially in cases where collapse occurs after multiple bounces from the 
AdS boundary.
This may have consequences for specific sets of initial data that are non-linearly stable against 
collapse (``islands of stability''). 
Namely, it is conceivable that the faster collapse we observe is an indication that these islands 
of stability are shrinking when spherical symmetry is broken. 
Systematic studies are required to investigate this point.

Our results further suggest that there exists a condition on the strength of the curvature 
deformation away from AdS in a given spatial volume, which separates data that collapse within 
one bounce of the condition being satisfied from data that will undergo further bounces.
Such a condition was obtained in the asymptotically flat case in spherical symmetry 
\cite{Christodoulou:1991yfa} and can be easily generalized to AdS \cite{PF}. 
It would be interesting to extend this analysis to the non-spherically symmetric setting.
We have also repeated parts of the present study in four spacetime dimensions and have obtained 
qualitatively similar results. 

We have broken spherical symmetry while preserving an $SO(3)$ symmetry in five dimensions. 
In this setting, we see that collapse is shifted away from the origin and leads to two distinct 
trapped regions on the axis. 
This can be understood as a focusing of energy density on the poles of the boundary $S^3$.
In the general case with no symmetry and arbitrary initial data, we expect this behavior to 
migrate away from the axis. 
A straightforward extension is to solve the momentum constraint equations along with the 
Hamiltonian constraint to generate configurations with non-zero total angular momentum. 
Based on results from perturbation theory, Ref. \cite{Dias:2017tjg} suggests that angular momentum 
may further enhance the non-linear instability of AdS (though see Ref. \cite{Choptuik:2017cyd}
for results of delayed collapse with the inclusion of angular momentum).
As we noted in the main text, it is not clear whether AdS black holes themselves are non-linearly 
stable, so completely general perturbations of AdS may not settle down.
We leave the question of stability of the black holes in this more general setting for a 
future work. 

%-------------------------------------------------------
% Acknowledgements
%-------------------------------------------------------
\PRLsection{Acknowledgements}%
We thank Andrzej Rostworowski, Luis Lehner, Alex Buchel, Frans Pretorius, and Maciej Maliborski 
for valuable discussions and comments.
We gratefully acknowledge the computer resources and the technical support provided by Princeton 
University and the Barcelona Supercomputing Center 
(FI-2016-3-0006 ``New frontiers in numerical general relativity"). 
H. B. and P. F. are supported by the European Research Council Grant No. 
ERC-2014-StG 639022-NewNGR. 
P. F. is also supported by a Royal Society University Research Fellowship (Grant No. UF140319). 
M. K. is supported by an STFC studentship. 
P. R. was supported in part by the Department of Energy, DOE award No. DE-SC0008132.
Simulations were run on the {\bf Perseus} cluster at Princeton University and the {\bf MareNostrum} 
cluster at the Barcelona Supercomputing Center. 

%-------------------------------------------------------
% Bibliography
%-------------------------------------------------------
\bibliography{arxiv_nov_07_2017}
\bibliographystyle{apsrev4-1}

%-------------------------------------------------------
% Supplemental Materials
%-------------------------------------------------------
\pagebreak

\makeatletter
\renewcommand{\thefigure}{S\@arabic\c@figure}
\makeatother

\makeatletter
\renewcommand{\theequation}{S\@arabic\c@equation}
\makeatother

\onecolumngrid

\begin{figure*}
{\large\bf Supplemental Material}
\end{figure*}

\twocolumngrid

\setcounter{figure}{0}
\setcounter{equation}{0}
\renewcommand{\theHfigure}{Supplement.\thefigure}
\renewcommand{\theHequation}{Supplement.\theequation}

%-------------------------------------------------------
% Evolution scheme
%-------------------------------------------------------
\PRLsection{Equations of motion}
Here we collect details of the scheme we use to evolve asymptotically AdS spacetimes.
We obtain initial data that are time symmetric and conformal to the pure AdS metric by solving the 
Hamiltonian constraint equation for an initial spatial metric sourced by a massless real scalar 
field. 
The spatial transition functions used in the construction of the scalar profiles in the main text 
are
\begin{eqnarray}\label{eqn:spatial_transitions}
f(\rho) &=& 1-h(\rho)\,, \nonumber \\
g(\rho) &=& 4 h(\rho) (1-h(\rho)) \nonumber \\
h(\rho) &=&
\left\{
\begin{array}{lll}
1                                   &,\, \rho \ge \rho_d \\
1-R^3 ( 6 R^2 - 15 R + 10)          &,\, \rho_d \ge \rho \ge \rho_c \\
0                                   &,\, \hbox{otherwise}
\end{array}
\right.. \nonumber \\
\end{eqnarray}
where $R(\rho) = (\rho_d-\rho)/(\rho_d-\rho_c)$.

We obtain subsequent times by solving the Einstein equations in generalized harmonic form with 
constraint damping
\begin{eqnarray}
&-& \frac{1}{2} g^{\alpha \beta} g_{\mu \nu, \alpha \beta} - 
{g^{\alpha \beta}}_{,(\mu} g_{\nu) \alpha, \beta} \nonumber \\
&-& H_{(\mu, \nu)} + H_\alpha\, {\Gamma^\alpha}_{\mu \nu} - {\Gamma^\alpha}_{\beta 
\mu} {\Gamma^\beta}_{\alpha \nu} \nonumber \\
&-& \kappa \left( 2\, n_{(\mu} C_{\nu)} - (1+P)\, g_{\mu \nu} \,n^\alpha 
C_\alpha \right) \nonumber \\
&=&   \frac{2}{3} \Lambda\, g_{\mu \nu} + 8\pi \left( T_{\mu \nu} - 
\frac{1}{3} {T^\alpha}_\alpha\, g_{\mu \nu} \right),
\end{eqnarray}
where $g_{\mu\nu}$ is the spacetime metric with Christoffel symbols ${\Gamma^\rho}_{\mu \nu}$,
$H_\mu$ are the source functions with constraints $C^{\mu} \equiv H^{\mu} - \square x^{\mu}$,
 $n^\mu$ is the unit (timelike) vector normal to the  $t=\textrm{const.}$ slices, and $\kappa$ and 
$P$ are constants. 
In the simulations presented here, we choose $\kappa=-10$ and $P=-1$. 
These equations are coupled to the equation of motion for a massless real scalar field, 
\begin{equation}
\Box\varphi=0\,,
\end{equation} 
with stress tensor
\begin{equation}
T_{\mu \nu} = \partial_\mu\varphi \, \partial_\nu \varphi - g_{\mu \nu} \left( \frac{1}{2} \, g^{\alpha \beta} \partial_{\alpha} \varphi \,\partial_{\beta} \varphi \right)\,.
\end{equation}

%-------------------------------------------------------
% Evolved Variables
%-------------------------------------------------------
%\PRLsection{Evolved Variables}
The evolved variables $\bar{g}_{\mu\nu}$ are constructed out of the full spacetime metric 
$g_{\mu\nu}$ and the pure AdS metric $\hat{g}_{\mu\nu}$ by
\begin{equation}\label{eqn:metric_falloff}
g_{\mu\nu} = \hat{g}_{\mu\nu} + (1-x^2-y^2) \bar{g}_{\mu\nu}\, ,
\end{equation}
where $x$ and $y$ are the Cartesian coordinates defined in the main text. For the invariant 
2-sphere piece of the metric, we define a single variable $\bar{g}_\psi$ so that
\begin{equation}\label{eqn:gbarpsi_definition}
\bar{g}_{\theta\theta}=\sin^2\theta\,\bar{g}_{\phi\phi} = y^2\,\bar{g}_\psi.
\end{equation}
The evolved variables $\bar{H}_\mu$ are similarly constructed out of the full spacetime source 
functions $H_\mu$ and the values $\hat{H}_\mu$ that they take on in pure AdS according to
\begin{equation}\label{eqn:sourcefunction_falloff}
H_\mu = \hat{H}_\mu + (1-x^2-y^2)^2 \bar{H}_\mu\,.
\end{equation}
Finally, the evolved variable for the scalar field $\bar{\varphi}$ is constructed out of a real 
scalar field $\varphi$ by
\begin{equation}\label{eqn:scalar_falloff}
\varphi = (1-x^2-y^2)^3 \bar{\varphi}. 
\end{equation}

%-------------------------------------------------------
% Gauge choice
%-------------------------------------------------------
%\PRLsection{Gauge choice}
The gauge choice is made by specifying the following form of the source functions:
\begin{eqnarray}
\bar{H}_\mu = \bar{H}^{(0)}_\mu\exp(-g_0) &+& F_\mu \left[ 1-\exp(g_0) \right] \nonumber \\
                                          &+& G_\mu \left[ 1-\exp(g_1) \right] \,,
\end{eqnarray}
where 
$\bar{H}^{(0)}_\mu=(\square x^{\mu}|_{t=0}-\hat{H}_\mu)/(1-x^2-y^2)^2$ 
are the initial values of the source functions, $F_\mu$ are components of the target gauge obtained 
by the procedure outlined in Ref. \cite{Bantilan:2012vu}
\begin{eqnarray}
F_t  &\equiv&  \frac{2 f_1}{\sqrt{x^2+y^2}}\left( x\,\bar{g}_{tx} + y\,\bar{g}_{ty} \right) \,, \nonumber \\
F_x  &\equiv& \frac{2 f_1}{\sqrt{x^2+y^2}}\left( x\,\bar{g}_{xx} + y\,\bar{g}_{xy} \right) \,, \nonumber \\
F_y  &\equiv& \frac{2 f_1}{\sqrt{x^2+y^2}}\left( x\,\bar{g}_{yy} + y\,\bar{g}_{xy} \right) \,,
\end{eqnarray}
and $G_\mu$ are the lapse damping terms used in Ref. \cite{Bantilan:2014sra}
\begin{eqnarray}
G_t  &\equiv& -c_1(1-f_2) \alpha\log\left(\textstyle{\frac{1}{\alpha}}\right)\,, \nonumber \\
G_x  &\equiv& -c_1(1-f_2)\frac{\beta_x}{\alpha} \,, \nonumber \\
G_y  &\equiv& -c_1(1-f_2)\frac{\beta_y}{\alpha}.
\end{eqnarray}
Here, $\alpha=-1/\sqrt{-g^{tt}}$ is the lapse function, $\beta_i = g_{ti}$ are the shift vector 
components, and
\begin{eqnarray}
g_0(t,\rho) &=&\frac{t^4}{\left( \xi_2 f_0(\rho) + \xi_1 [1-f_0(\rho)] \right)^4}\,, \nonumber \\
g_1(t,\rho) &=&\frac{t^4}{\xi_3^4}\,, \nonumber \\
f_k(\rho) &=&
\left\{
\begin{array}{lll}
1                                &,\, \rho \ge \rho_{2k+2} \\
1-R_k^3 ( 6 R_k^2 - 15 R_k + 10) &,\, \rho_{2k+2} \ge \rho \ge \rho_{2k+1} \\
0                                &,\, \hbox{otherwise}
\end{array}
\right., \nonumber \\
\end{eqnarray}
where $R_k(\rho) = (\rho_{2k+2} - \rho)/(\rho_{2k+2} - \rho_{2k+1})$, and 
$\rho_1$, $\rho_2$, $\rho_3$, $\rho_4$, $\rho_5$, $\rho_6$, $\xi_1$, $\xi_2$, $\xi_3$, $c_1$ 
are constants. 
On a typical run, we set 
$\rho_1=0.0$, $\rho_2=0.95$, $\rho_3=0.05$, $\rho_4=0.95$, $\rho_5=0.0$, $\rho_6=0.5$, $\xi_1=0.1$, 
$\xi_2=0.0025$, $\xi_3=0.1$, $c_1=20.0$.

%-------------------------------------------------------
% Boundary conditions
%-------------------------------------------------------
%\PRLsection{Boundary conditions}
We set Dirichlet boundary conditions at spatial infinity for the metric, source functions, and 
scalar field:
\begin{eqnarray}\label{eqn:bcs}
\left. \bar{g}_{\mu\nu} \right|_{\rho=1}        &=& 0 \,, \nonumber \\
\left. \bar{H}_{\mu} \right|_{\rho=1}           &=& 0 \,,\nonumber \\
\left. \bar{\varphi} \right|_{\rho=1}           &=& 0 \,. 
\end{eqnarray}
In practice, because of our Cartesian grid, $\rho=1$ does not necessarily lie on a grid point aside 
from the exceptional points $x=\pm 1,y=0$ and $x=0,y=1$. 
We thus implement \eqref{eqn:bcs} on the grid points that are closest to $\rho=1$, i.e., points 
that are at most one grid point away from $\rho=1$. 
These near-boundary points are set using \eqref{eqn:bcs} at $\rho=1$ and data at the point that are 
one grid point interior to the point in question by linear interpolation in the Cartesian 
directions: along $x$ for near-boundary points with $|x|>y$, and along $y$ for near-boundary 
points with $|x|<y$. 
 
On the symmetry axis $y=0$ we impose regularity conditions on all evolved variables. 
These are:
\begin{eqnarray}\label{eqn:metric_axireg}
\left. \partial_y \bar{g}_{tt} \right|_{y=0}      &=& 0\,, \nonumber \\
\left. \partial_y \bar{g}_{tx} \right|_{y=0}      &=& 0 \,, \nonumber \\
\left. \partial_y \bar{g}_{xx} \right|_{y=0}      &=& 0 \,,\nonumber \\
\left. \partial_y \bar{g}_{yy} \right|_{y=0}      &=& 0 \,,\nonumber \\
\left. \partial_y \bar{g}_{\psi} \right|_{y=0}      &=& 0 \,,\nonumber \\
\left. \bar{g}_{xy} \right|_{y=0}                 &=& 0 \,,\nonumber \\
\left. \bar{g}_{ty} \right|_{y=0}                 &=& 0 \,,\nonumber \\
\left. \partial_y \bar{H}_{t} \right|_{y=0}         &=& 0 \,,\nonumber \\
\left. \partial_y \bar{H}_{x} \right|_{y=0}         &=& 0 \,,\nonumber \\
\left. \bar{H}_{y} \right|_{y=0}         &=& 0 \,,\nonumber \\
\left. \partial_y \bar{\varphi} \right|_{y=0}         &=& 0\,. 
\end{eqnarray}
In addition, we demand that there be no conical singularities at $y=0$, which amounts to requiring 
that  
$\left. \bar{g}_{yy} \right|_{y=0} = \left. \bar{g}_{\psi} \right|_{y=0}$. 
We impose this condition on $\bar{g}_{yy}$ at $y=0$ instead of the corresponding regularity 
condition for $\bar{g}_{yy}$ in \eqref{eqn:metric_axireg}.

%-------------------------------------------------------
% CFT Stress Tensor
%-------------------------------------------------------
\PRLsection{CFT Stress Tensor}
\begin{figure*}[t]
        \includegraphics[height=0.22\linewidth]{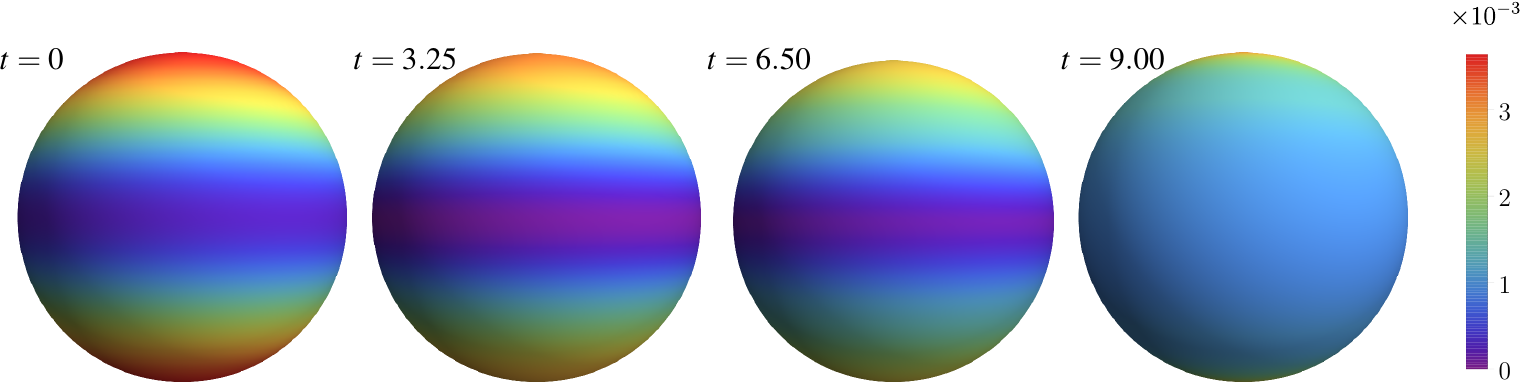}
                        {\caption{
                            Snapshots of the energy density on the boundary of global AdS at different global times $t$ for the simulation with $A=0$, $B=0.0087$ and a total mass $M=0.021$. The energy density starts off peaked at the poles, and oscillates back and forth between the poles and the equator in a quasi-periodic fashion. Gravitational collapse into two black holes on the axis at $t=9.04$ is seen as a localization of the energy density at the poles of the boundary $S^3$.
                        }\label{fig:sphere}}
\end{figure*}
Here we briefly spell out how we extract the expectation value 
$\left< T_{\mu \nu} \right>_{\text{CFT}}$ of the CFT stress energy tensor from the asymptotic 
behavior of the metric by
\begin{equation}\label{eqn:cftsetexpectation}
\left< T_{\mu \nu} \right>_{\text{CFT}}=\underset{\rho \rightarrow 1}{\lim}{\frac{1}{(1-\rho)^2} {}^{(\rho)} \! T_{\mu \nu}}.
\end{equation}
${}^{(\rho)} T_{\mu \nu}$ is the Brown-York quasi-local stress tensor \cite{Brown:1992br}
defined on a $\rho={\rm const.}$ surface, given by Ref. \cite{Balasubramanian:1999re}
\begin{equation}\label{eqn:quasiset}
{}^{(\rho)} T^0_{\mu \nu} = \frac{1}{8 \pi} \left( {}^{(\rho)}\Theta_{\mu \nu} - {}^{(\rho)}\Theta \Sigma_{\mu \nu} 
- \frac{3}{L} \Sigma_{\mu \nu} + {}^{(\rho)} G_{\mu \nu} \frac{L}{2} \right).
\end{equation}
Here, ${}^{(\rho)}\Theta_{\mu \nu} = -{\Sigma ^\alpha}_\mu {\Sigma ^\beta}_\nu \nabla_{(\alpha} S_{\beta)}$
is the extrinsic curvature of the $\rho={\rm const.}$ surface, $S^\mu$ is a space-like outward 
pointing unit vector normal and $\Sigma_{\mu \nu}\equiv g_{\mu\nu} - S_\mu S_\nu $ is the induced 
4-metric on the surface, $\nabla_\alpha$ is the covariant derivative operator and 
${}^{(\rho)} G_{\mu \nu}$ is the Einstein tensor associated with $\Sigma_{\mu \nu}$. 
The last two terms in (\ref{eqn:quasiset}) are counterterms designed to cancel the divergent 
boundary behavior of the first two terms of~(\ref{eqn:quasiset}) evaluated in pure AdS$_5$. 
The stress tensor (\ref{eqn:quasiset}) is non-vanishing even in pure global AdS$_5$: 
the CFT is defined on the boundary which has topology $\mathbb{R} \times S^3$, and so can have a
non-vanishing Casimir vacuum energy. Since this is a constant non-dynamical contribution, 
we consider it as part of our background vacuum and simply subtract it from (\ref{eqn:quasiset}). 

The conserved mass $M$ of the spacetime is computed from the quasi-local stress tensor 
\eqref{eqn:quasiset} as follows: we take a spatial $t={\rm const.}$ slice of the 
$\rho={\rm const.}$ surface, with induced 3-metric $\sigma_{\mu \nu}$, lapse $N$ and shift $N^i$ 
such that
$\Sigma_{\mu \nu} dx^\mu dx^\nu = -N^2 dt^2 + \sigma_{ij}(dx^i + N^i dt)(dx^j +N^j dt)$ and
we compute
\begin{equation}\label{eqn:adsmass}
M = \underset{\rho \rightarrow 1}{\lim} \int_{\Sigma} d^3 x \sqrt{\sigma} N ( {}^{(\rho)} T_{\mu \nu} u^\mu u^\nu )\,,
\end{equation}
where $u^\mu$ is the time-like unit vector normal to $t={\rm const.}$ 
In particular, because of background subtraction described above, we recover a vanishing mass for 
pure AdS.

The energy density in terms of the CFT stress tensor components $T_{\mu\nu} \equiv \left< T_{\mu \nu} \right>_{\text{CFT}}$ is
\begin{eqnarray}\label{eqn:hydroextraction}
\epsilon &=& \frac{1}{2} \left(T_{tt} -T_{\chi\chi} + \sqrt{(T_{\chi\chi} + 2 T_{t\chi}+T_{tt}) (T_{\chi\chi} - 2 T_{t\chi}+T_{tt})} \right). \nonumber \\
\end{eqnarray}

The evolution of the energy density as a function of the global time $t$ for the simulation with 
$A=0$, $B=0.0087$ and a total mass $M=0.021$ is shown in Fig.~\ref{fig:sphere}. 
The energy density exhibits quasi-periodic behavior until it localizes at the poles. 
The latter corresponds to the formation of two black holes on the axis in the bulk.  

\begin{figure*}[t]
  \includegraphics[width=0.48\linewidth]{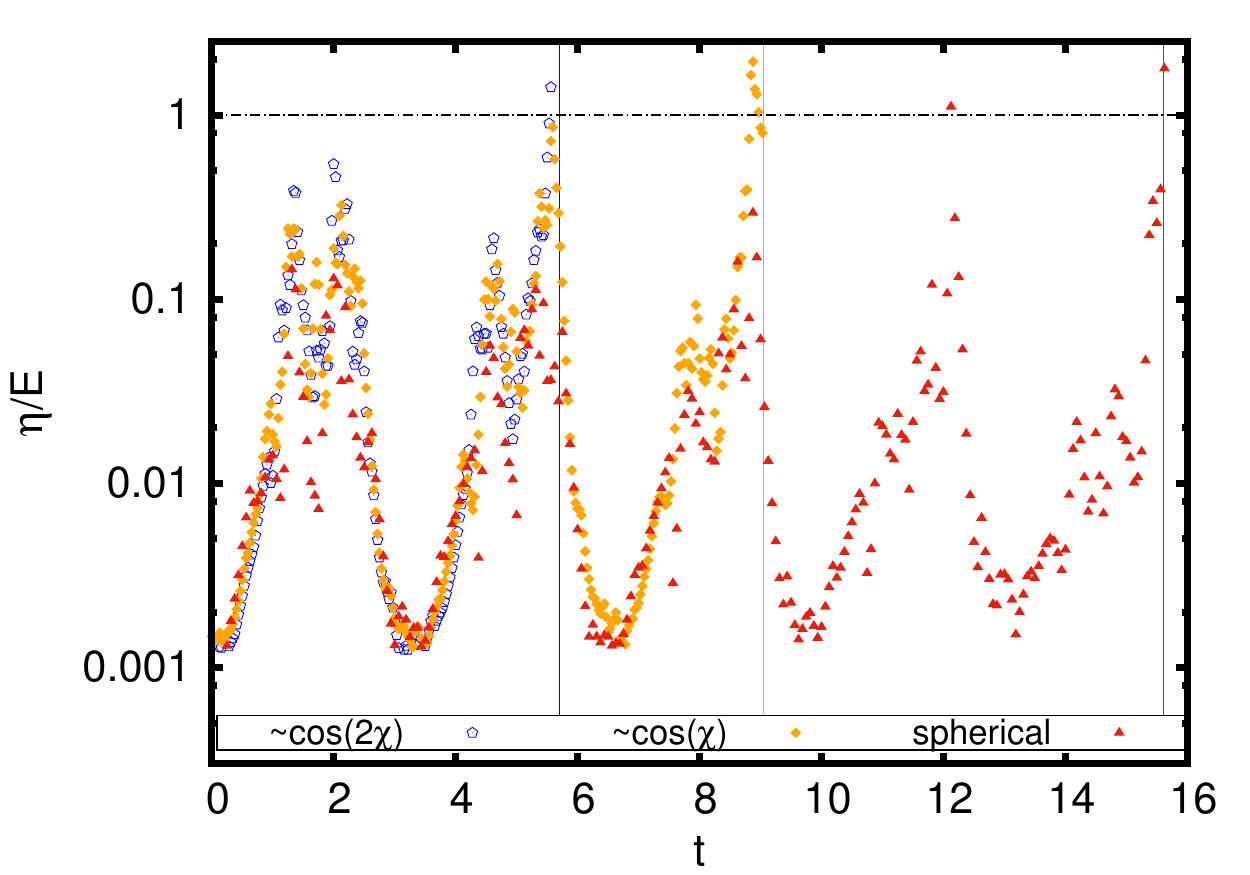}
  \hfill
  \includegraphics[width=0.48\linewidth]{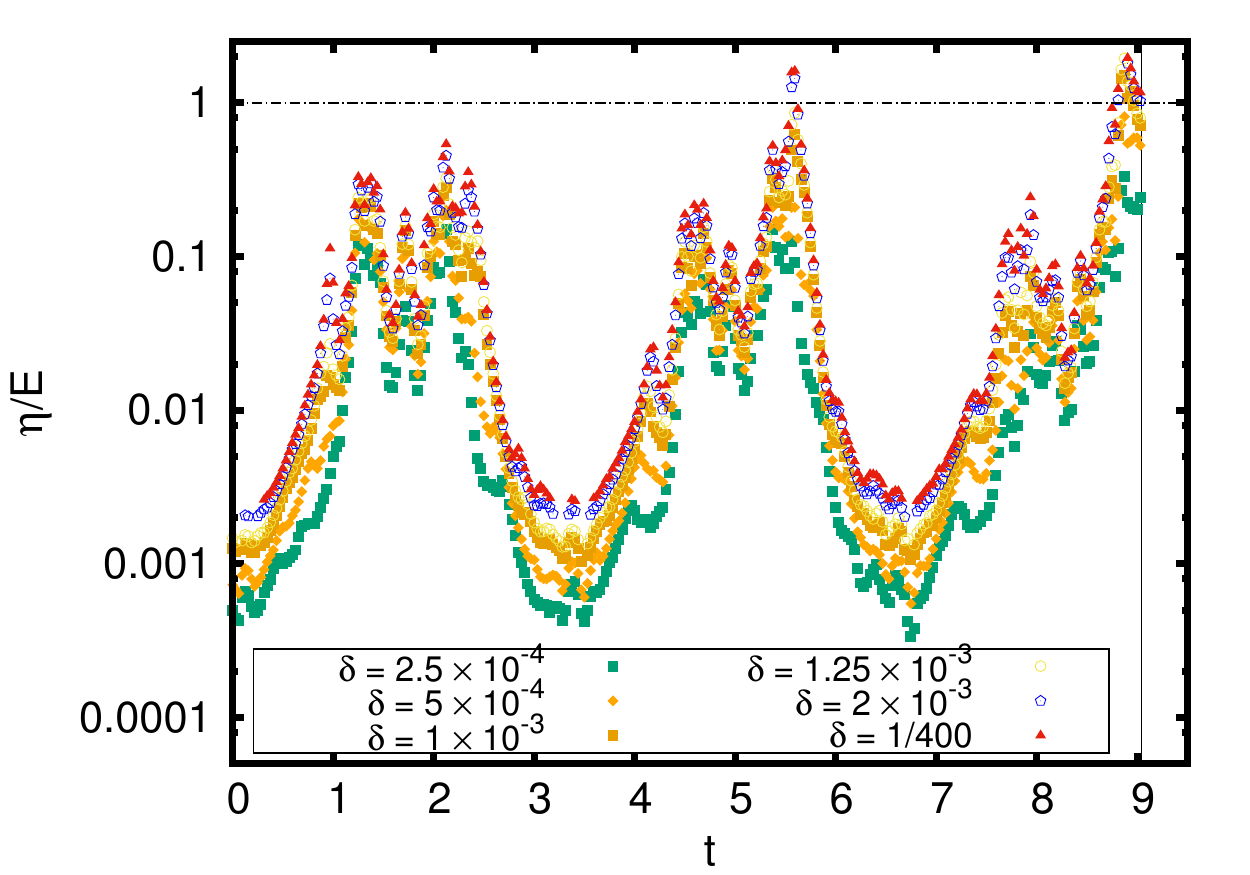}
                        {\caption{
                          \textit{Left}: Comparison of three simulations with the same total mass of $M=0.021$ but using initial scalar profiles with different $\chi$ dependence.
                          \textit{Right}: The effect of changing the threshold $\delta$ defined in the main text, for the simulation with $A=0$, $B=0.0087$ and a total mass 
$M=0.021$.
                        }\label{fig:harmonics_thresholds}}
\end{figure*}

%-------------------------------------------------------
% Higher Harmonics and Different Thresholds
%-------------------------------------------------------
\PRLsection{Higher Harmonics and Different Thresholds}
Fig.~\ref{fig:harmonics_thresholds} (\textit{left}) shows three simulations of spacetimes with the 
same total mass, but deformed from pure AdS by spatial profiles with different $\chi$ dependence. 
The initial data deformed with the higher harmonics collapse earlier. 
Fig.~\ref{fig:harmonics_thresholds} (\textit{right}) shows the effect of changing the threshold 
$\delta$ for the simulation with $A=0$, $B=0.0087$ and a total mass $M=0.021$. 
For smaller values of $\delta$, i.e., for a less strict condition and thus a larger integration
region, the maximum value of $\eta/E$ decreases, indicating that these points are associated with 
configurations where the Kretschmann scalar is very sharply peaked in smaller regions. 
Changing $\delta$ does not affect the time $\eta/E$ attains this maximum.

%-------------------------------------------------------
% Numerical tests
%-------------------------------------------------------
\PRLsection{Numerical tests}
We check our solutions using a pair of standard convergence tests. To check for stability and 
consistency, we compute the rate of convergence $Q(t,x,y)$ at each point on the grid for a given 
evolved variable
\begin{equation}\label{eq:qconv}
Q(t,x,y)=\frac{1}{\ln(2)}\ln\left( \frac{f_{4h}(t,x,y)-f_{2h}(t,x,y)}{f_{2h}(t,x,y)-f_{h}(t,x,y)} \right).
\end{equation}
Here, $f_h$ denotes one of $\bar{g}_{\mu\nu},\bar{H}_\mu,\bar{\varphi}$ from a simulation with 
mesh spacing $h=\Delta x = \Delta y$. 
We use second-order accurate finite difference stencils, with $2:1$ refinement in $h$ between
successive resolutions. 
Since we keep the CFL factor constant at $\lambda=0.2$, the time step is decreased by the same 
ratio. 
We thus expect $Q$ to asymptote to $Q=2$ in the limit $h\rightarrow0$.

\begin{figure*}[t!]
  \includegraphics[width=0.48\linewidth]{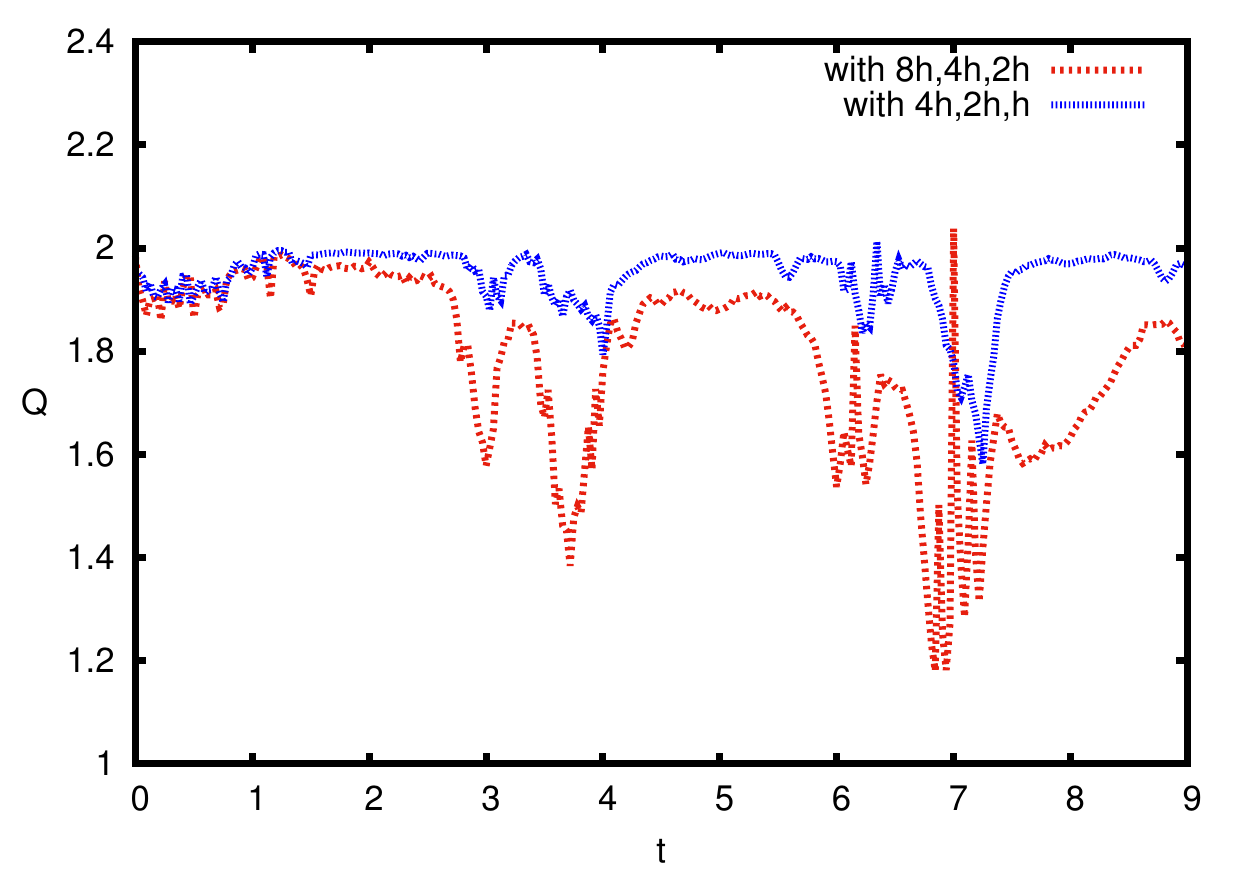}
  \hfill
  \includegraphics[width=0.48\linewidth]{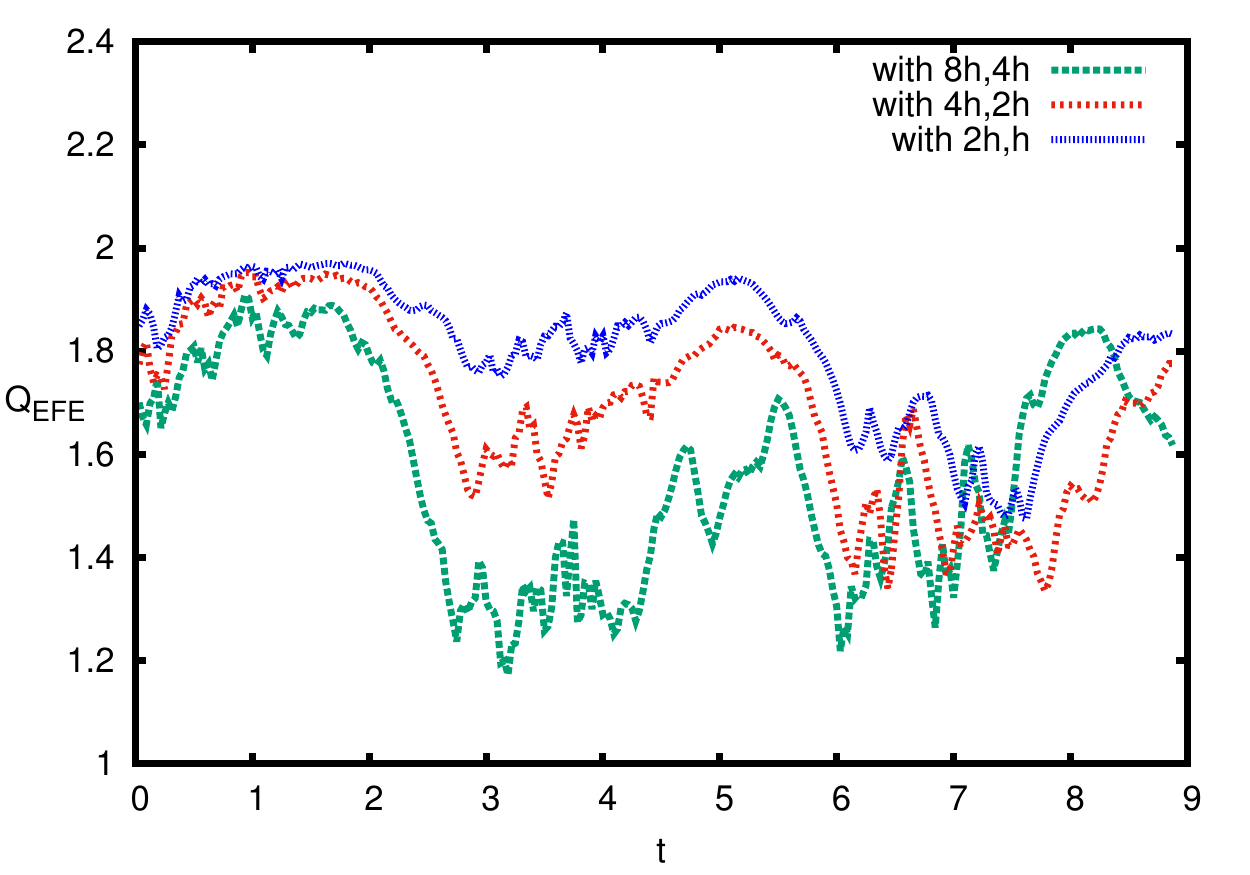}
                        {\caption{
                          \textit{Left}: Convergence factors for the $\bar{g}_{xx}$ metric variable from a simulation with $A=0$, $B=0.0087$ and a total mass $M=0.021$.
                          \textit{Right}: Convergence factors for the independent residual from the same simulation. In each panel, the $L^2$ norm of the convergence factors is taken over the entire 
grid.
                        }\label{fig:convergence_tests}}
\end{figure*}

To check that our numerical solutions are converging to a solution of the Einstein equations, we 
compute an independent residual by taking the numerical solution and substituting it into a 
discretized version of
$G_{\mu\nu} + \Lambda g_{\mu\nu} - 8\pi T_{\mu\nu}$. 
Since the numerical solution was found solving the generalized harmonic form of the Einstein 
equations, we expect the independent residual to only be due to numerical truncation error and 
thus converge to zero. 
We can compute a convergence factor for the independent residual by
\begin{equation}\label{eq:qires}
Q_{EFE}(t,x^i)=\frac{1}{\ln(2)}\ln\left( \frac{f_{2h}(t,x^i)}{f_{h}(t,x^i)} \right).
\end{equation}
Here, $f_h$ denotes a component of
$G_{\mu\nu} + \Lambda g_{\mu\nu} - 8\pi T_{\mu\nu}$.
Again, given our second-order accurate finite difference stencils and with $2:1$ refinement in $h$ 
between successive resolutions, we expect $Q$ to approach $Q=2$ as $h\rightarrow0$.

\end{document}